\documentclass[preprint]{emulateapj} 

\usepackage{epsfig}
\usepackage{graphicx}

\shorttitle{On-sky speckle nulling with SCExAO}
\shortauthors{Martinache et al}

\begin{document}

\title{on-sky speckle nulling demonstration at small angular
  separation with SCExAO}

\author{
  Frantz Martinache\altaffilmark{1, 2}, 
  Olivier Guyon\altaffilmark{1, 3}, 
  Nemanja Jovanovic\altaffilmark{1}, 
  Christophe Clergeon\altaffilmark{1}, 
  Garima Singh\altaffilmark{1},
  Tomoyuki Kudo\altaffilmark{1}, 
  Thayne Currie\altaffilmark{8},
  Christian Thalmann\altaffilmark{6, 7}, 
  Michael McElwain\altaffilmark{4}, 
  Motohide Tamura\altaffilmark{7}}

\altaffiltext{1}{National Astronomical Observatory of Japan, 
  Subaru Telescope, Hilo, HI 96720, USA}
\altaffiltext{2}{Laboratoire Lagrange, UMR7293, Universit\'e de Nice
  Sophia-Antipolis, CNRS, Observatoire de la C\^ote d’Azur, Bd. de
  l’Observatoire, 06304 Nice, France}
\altaffiltext{3}{Stewart Observatory, University of Arizona,
  Tucson, AZ 85721, USA}
\altaffiltext{4}{Goddard Space Flight Center, Greenbelt, MD, USA}
\altaffiltext{5}{Astronomical Institute ``Anton Pannekoek'',
  University of Amsterdam, Science Park 904, 1098 XH Amsterdam, The
  Netherlands}
\altaffiltext{6}{Department for Astronomy, ETH Zurich, CH-8093 Zurich,
Switzerland}
\altaffiltext{7}{National Astronomical Observatory of Japan, 2-21-1
  Osawa, Mitaka, Tokyo 181-8588, Japan}
\altaffiltext{8}{Department of Astronomy and Astrophysics, University
  of Toronto, Canada}
\email{frantz@naoj.org}

\begin{abstract}
This paper presents the first on-sky demonstration of speckle nulling,
which was achieved at the Subaru Telescope in the context of the
Subaru Coronagraphic Extreme Adaptive Optics (SCExAO) Project.
Despite the absence of a high-order high-bandwidth closed-loop AO
system, observations conducted with SCExAO show that even in
poor-to-moderate observing conditions, speckle nulling can be used to
suppress static and slow speckles even in the presence of a brighter
dynamic speckle halo, suggesting that more advanced high-contrast
imaging algorithms developed in the laboratory can be applied
to ground-based systems.
\end{abstract}


\section{Introduction}

The detection of high-contrast features such as disks and companions
in diffraction-limited images relies for the most part on the
characterization and calibration of diffraction effects, whether
static (induced by the geometry of the pupil and the aberrations in
the optics), quasi-static (telescope pointing and slowly variable
instrument alignment) or dynamic (atmosphere-induced).
For ground-based observations, the best performance comes from
a combination of adaptive optics (AO), which address the dominant
atmospheric term, turning the seeing-limited image into a
diffraction-limited one, and post-processing techniques based on
differential imaging. These include spectral, polarimetric or angular
differential imaging (ADI), the first two relying upon some properties
of the target, such as the presence of specific spectral features or a
polarimetric signature.

ADI \citep{2006ApJ...641..556M, 2007ApJ...660..770L} provides the
means to calibrate the static and quasi-static features of the PSF, by
observing a target with an alt-azimuthal telescope, for which the sky
rotates relative to the telescope pupil. If enough field rotation
occurs over a time that is less than the characteristic quasi-static
aberration time-scale, the diversity between the orientation of the
sky and the quasi-static point spread function (PSF) allows the
calibration of the diffraction pattern, leading to greatly enhanced
detection limits.

The calibration requires that the angular rotation induces sufficient
local linear displacement of the genuine features such as faint
companions relative to the quasi-static PSF
\citep{2006ApJ...641..556M}, in order to avoid self-subtraction.
While highly efficient at angular separations greater than 1\arcsec,
it is difficult to benefit from this technique at angular separations
smaller than 0.5 \arcsec. 
For addressing such small angular separations, two approaches are
possible: interferometric calibration and additional active wavefront
control.

Interferometric calibration of diffraction-limited images is a
nascent field, that inherits from the ideas and techniques of sparse
aperture masking (SAM) a.k.a. non-redundant masking (NRM)
interferometry with AO \citep{2006SPIE.6272E.103T}.
The fundamental idea behind these techniques is that although the
content of images may be corrupted by residual aberrations, it is
possible to construct a sub-set of information that is independent 
from these aberrations. One possible such sub-set of information is
the kernel-phase \citep{2010ApJ...724..464M}, which just like sparse
aperture masking using closure phase, goes as far as enabling the direct
detection of high contrast features beyond the diffraction limit, a
regime referred to as super-resolution \citep{2013ApJ...767..110P}.
The fundamental advantage of this approach is that it requires no
additional differential ``trick'' like ADI, although it is for now
restricted to well-corrected AO images.
Recent improvements of the technique, such as the statistically
independent kernel-phase and better calibration procedures described
by \citet{2013MNRAS.433.1718I} show promise of performance that
compares to what is currently achieved by ADI at small angular
separations.

\begin{figure*}[htb!]
\plotone{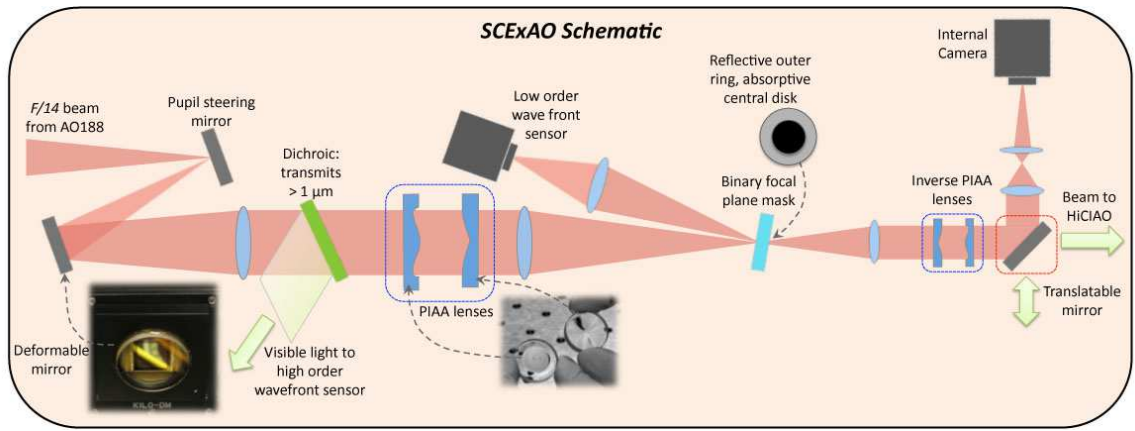}
\caption{
Schematic representation of the optics inside the SCExAO IR
coronagraph used to produce the results presented in this paper.
}
\label{f:layout} 
\end{figure*}

The alternative approach to the passive calibration by interferometry
is to employ additional wavefront control to modulate the diffraction
at small angular separation, and create the diversity that enables the
separation of genuine structure from diffraction features in the
image. This concept, which can be referred to as coherence differential
imaging (CDI), is very relevant to the upcoming generation of so
called extreme AO (XAO) projects:
Palm-3000/P1640 at Palomar \citep{2012SPIE.8447E..20O}, GPI at Gemini
\citep{2008SPIE.7015E..31M}, SPHERE at VLT \citep{2010SPIE.7736E..13S}
and SCExAO at the Subaru Telescope \citep{2011SPIE.8151E..22M}.

Among these projects, the Subaru Coronagraphic Extreme AO (SCExAO)
project is unique with its incremental deployment schedule that
enabled the on-sky testing of fundamental features of the instrument
(Jovanovic et al, in prep) prior to completion of its closed-loop XAO
system \citep{2013AAS...22130504C}. In this paper, we present the
first convincing demonstration of on-sky closed-loop diffraction
modulation, after an upstream AO correction is performed, providing a
Strehl on the order of 20\%. While the demonstrated gain is moderate
and there is room for improvement, the paper nevertheless demonstrates
that it is possible to control the features of a coronagraphic PSF
near the inner working angle of the coronagraph, using the speckle
nulling approach introduced in previous work.

\section{SCExAO before XAO}

What is usually referred to as an XAO system is a unit capable of
producing a very stable and high-quality PSF, characterized by a
Strehl ratio on the order of 90\%. This very high level of Strehl,
corresponding to residual wavefront errors on the order of
$\lambda/20$ (where $\lambda$ is the wavelength of observation), is
indeed required for the high-contrast technique of coronagraphy, that
aims to block the light of an on-axis bright source and reveal its
circumstellar environment, in particular disk and/or faint
companions.

Particularly in monochromatic light, it is possible to design
coronagraphs that provide near-perfect extinction, and an inner
working angle (IWA) that approaches the limit of diffraction
($\lambda/D$), where $D$ represents the diameter of the telescope
aperture. Examples of such coronagraphs include the vortex
\citep{2010ApJ...709...53M}, and the PIAA \citep{2003A&A...404..379G},
both implemented on SCExAO. 
In practice, however, their performance is entirely dictated by the
residual wavefront errors. Under poor Strehl conditions, the
coronagraph does little more than allow the exposure time to be
increased before saturation of the detector.

SCExAO does not currently support a high-order closed-loop AO system,
expected to be first commissioned in the fall of 2014. In this first
phase of the project, SCExAO therefore relies entirely on the 
up-stream Subaru Telescope facility AO system called AO188
\citep{2010SPIE.7736E.122M}.
AO188 and the coronagraphic imager HiCIAO \citep{2008SPIE.7014E..42H}
form the workhorse of the Strategic Exploration of Exoplanets and
Disks with Subaru (SEEDS) observing campaign
\citep{2009AIPC.1158...11T}.
SCExAO was conceived as a replacement upgrade for the current
fore-optics used by HiCIAO, that should include a small IWA PIAA-based
coronagraph as well as additional wavefront manipulation capability.

While SCExAO therefore does not currently qualify as an XAO system, it
nevertheless already implements a 1-k actuator deformable mirror (DM).
We have recently shown, using a calibration source in a stable
laboratory environment that the DM and the PIAA coronagraph used
together can produce a high-contrast region in the field-of-view, with
a 2.2 $\lambda/D$ IWA \citep{2012PASP..124.1288M}, with a simple
iterative speckle-nulling loop. 
Using the same approach on-sky, relying on AO188 to produce a 20\%
Strehl ratio PSF, we demonstrate that it is possible to use speckle
nulling to drive down the brightness of the diffraction features at
small angular separation. We also show that the DM provides a flexible
calibration tool that notably enables astrometric calibration as well
as direct measurements of the contrast in images.

\section{On-sky coronagraphic calibration}
\label{sec:cal}

Figure \ref{f:layout} provides a schematic overview of the optical
layout of the near-IR arm of SCExAO, as it was used on the observing
night of November 12, 2012.
The upstream AO188 system feeds the instrument with a partially
corrected f/14 convergent beam, intercepted by a first tip-tilt
controlled mirror located in an image plane and mounted on a focusing
stage, and further steered by a second tip-tilt controlled
mirror. This combination provides full control of the origin and
incidence of the beam entering the coronagraph.
The other important feature, is the focal plane mask, reflecting the
light of any on-axis source onto a camera used to sense the pointing,
a sub-system called the coronagraphic low-order wavefront sensor
(CLOWFS), as described by \citet{2009ApJ...693...75G} and
\citet{2011PASP..123.1434V}.

In addition, the system also comprises the PIAA remapping optics
\citep{2005ApJ...622..744G} located before the focal plane mask and
their inverse counterpart located after the focal plane mask. 
Recent work \citep{2012PASP..124.1288M} showed that when used
together, the presence of these two sets of remapping optics can be
ignored by the focal plane based wavefront control algorithm. 
Reflecting upon this conclusion, this discussion will ignore the
remapping optics, which for the results presented in this paper, are
always in the beam.

One of the two tip-tilt mirrors that steers the beam entering the
coronagraph is the 1024-actuator DM itself, which while not exactly
located in a pupil plane, can nevertheless be used to
introduce/compensate for diffraction features in the coronagraphic
image.
The square grid geometry of the actuators underneath the DM membrane
makes it particularly well suited to generate pairs of speckles, that
interfere with already present diffraction features in the
coronagraphic image.
The total number of actuators across the telescope pupil footprint
imposes the size of the region of the coronagraphic image that can be
probed by the DM, which will be referred to as the control region.
As explained by \citet{2012PASP..124.1288M}, the control region with
this current implementation fits within a 27.2 x 24.8 $\lambda/D$
rectangular box.
It is possible to generate a pair of speckles centered around the
optical axis within this region, by applying a sinusoidal displacement
map on the DM.

\begin{figure}
\center{\resizebox*{0.5\columnwidth}{!}{
    \includegraphics{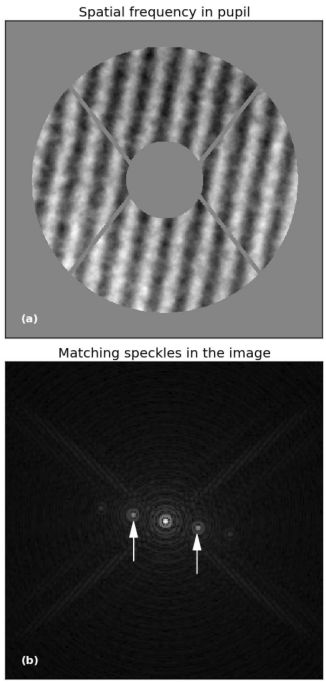}}}
\caption{
Illustration of the relation between sinusoidal modulation of the DM
shape in the pupil and the resulting diffraction pattern in the
image. In panel (a), one can approximately count 10 cycles of the
sinusoid across the entire pupil. In the image (panel (b)), this
modulation of the DM creates a pair of speckles (pointed by white
arrows) located 10 $\lambda/D$ away from the center of the field.
These additional speckles interfere with the already present
diffraction features such as rings, spikes and phase
defects. They however do not interfere with incoherent structures like
planets and disks.
}
\label{f:modul} 
\end{figure}

Each such sinusoidal displacement map on the DM is characterized by
four numbers: two cartesian coordinates of spatial frequency 
$(k_x, k_y)$, one amplitude $\alpha$ (in radians) limited by the
stroke of the DM and a phase $\varphi$, that can take any value
between 0 and 2$\pi$.
When applying a sinusoidal displacement map, the DM acts like a
diffraction grating that creates off-axis copies of the on-axis bright
source. Figure \ref{f:modul} illustrates the relation between DM
modulation and speckles in the image.
The larger the spatial frequency parameters, the faster the sinusoidal
modulation of the DM surface and the further the corresponding pair of
speckles lies away from the center of the control region. The larger
the amplitude of the modulation, the brighter the resulting
speckles.
The phase does not change the intensity of the speckles, unless they
interfere with other coherent features in the image.
Sinusoidal modulations of distinct properties can be added to the DM
as long as the total requested range of displacement does not go
beyond the DM stroke, creating in turn multiple pairs of speckles in
the image.

The ability to produce symmetric pairs of speckles of controllable
positions and contrast is a very convenient feature that facilitates
otherwise difficult calibration procedures when observing with a
coronagraph.
Indeed, unlike classical imaging techniques and interferometric
methods for which absolute control of the pointing is not a strict
requirement since the central star remains visible, the image of the
star needs to be precisely driven and stabilized on the axis of the
coronagraph for an efficient suppression.

\citet{2006ApJ...647..620S} astutely proposed to address
this issue by inserting a reticulate grid of wires in the pupil plane,
that produces a pre-defined periodic diffraction pattern in the image.
The grid geometry of deformable mirrors such as the 1k-actuator MEMS
DM used in SCExAO can advantageously be used as an adaptive version of
this reticulate grid.
Just like described by Sivaramakrishnan \& Oppenheimer, the added
satellite speckles provide a very reliable way to position the star
behind the focal plane mask during target acqusition. Once in the
ideal position, the low-order wavefront sensor
\citep{2009ApJ...693...75G} keeps the image in a fixed position
relative to the mask.
During the post-processing stage, in the case non-coherent
structures such as companions are identified in the image, the
satellite speckles offer a reliable astrometric reference.

\begin{figure}
\center{\resizebox*{0.5\columnwidth}{!}{
    \includegraphics{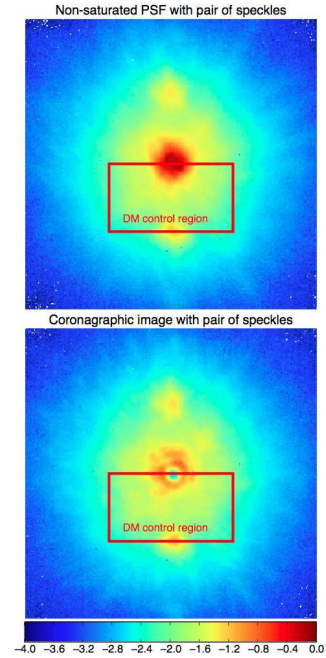}}}
\caption{
  Comparison of a non-coronagraphic (top panel) and a coronagraphic
  (bottom panel) image acquired by HiCIAO behind SCExAO, with an added
  pair of speckles at max angular separation (log contrast scale).
  The red box highlights the area controlled by the DM during the
  experiments reported in this paper.
  Added speckles like these enable the calibration of the image in
  contrast,   even if the central star is blocked by the coronagraph.
}
\label{f:contr} 
\end{figure}

In addition to offering astrometric and pointing aids
\citep{2013ApJ...779..153H}, the generated speckles can be used for
contrast calibration purposes. This can efficiently be achieved by an
off-line pre-determination of the satellite speckle contrast relative
to the central PSF as a function of spatial frequency and amplitude.
Figure \ref{f:contr} shows one example of such on-sky contrast
calibrated images, recorded by the coronagraphic imager HiCIAO fed 
light by SCExAO.
The top panel of Fig. \ref{f:contr} shows a non-saturated low-exposure
time image with the SCExAO focal plane mask out of the way. A
sinusoidal modulation of the SCExAO DM of known characteristics
creates a pair of speckles of controlled contrast.
The bottom panel shows an image acquired with the same exposure time,
this time with the SCExAO focal plane mask placed on-axis. Except at
small angular separation, one can observe very little impact of the
focal plane mask on the halo PSF features.

\section{On-sky speckle nulling}

\citet{2012PASP..124.1288M} showed, using SCExAO under stable
laboratory conditions, that it is possible to actuate the DM before
the PIAA coronagraph to create diffraction features that destructively
interfere with speckles in the field, creating a uniform high contrast
region, often referred to as a ``dark hole'' in the image.
Here, we apply the same iterative speckle nulling algorithm during
actual on-sky observations. The major difference with the laboratory scenario
is that the static (or quasi-static) features of the PSF probed by the
speckle nulling algorithm are buried underneath a very-strong,
fast-varying dynamic component, due to the incomplete AO correction,
in the absence of a properly closed XAO loop.

The reported performance of ADI calibration suggests that
high-contrast detection limits are set by long-lived aberrations, with
a characteristic time-scale $\sim$1 hour or less.
In this experiment, the science detector (HiCIAO) and the
speckle-nulling camera cannot simultaneously observe: a retractable
mirror sends the light either way.
To be useful, the speckle nulling loop must therefore converge within
a time-scale on the order of 15-20 minutes, so that sufficient time is
left for the acquisition of frames by the (slow) science detector,
before the quasi-static aberrations start changing again.

\subsection{Technical constraints}
This on-sky speckle nulling experiment was performed using an uncooled
InAsGa type detector (Xeva XS 1.7-320 sold by Xenics), and a
H-broadband filter (1.6 $\mu$m).
Such cameras exhibit frame rates up to $\sim$100 Hz that make them
appealing for NIR wavefront control applications however, they also
have fairly high readout noise ($>$300 e$^-$) as well as high dark
current.
In the high-gain setting, the dark current averages 10000 counts in 20
ms exposure (almost the entire usable dynamic range): coronagraphic
observations with these cameras are therefore limited to very bright
stars only.

The images showed here were obtained while observing the bright K-type
star Pollux ($V=1.15, R=0.6, H=-0.845$). For this object, the best
compromise was to use a 10-ms exposure time in low-gain mode, with a
median dark current level $\sim$4200.
While the speckle nulling loop is running, the camera continuously
acquires 10-ms exposures at the frame rate of 30 Hz.

In a given image, up to $n$ speckles are identified within the
control region and their positions marked, relative to the central
source, hidden by the coronagraph, but nevertheless located by the
satellite speckles as described in Section \ref{sec:cal}.
The first step is to identify the two-component spatial frequency
$(k_x,k_y)$ of the DM that corresponds to each of the speckles
identified inside the control region (cf. Fig. \ref{f:modul}).
The amplitude $\alpha_0$ of this spatial frequency is estimated
from the speckle brightness, proportional to the square of the
amplitude $\alpha_0$.
The only remaining unknown is the phase $\varphi$ of the speckle,
that can take any value between 0 and 2$\pi$.

To determine this unknown, one modulates the DM with a sinusoidal
function so as to generate a speckle probe of constant amplitude
$\alpha_0$, but whose phase is varied from 0 to 2$\pi$ radians from
image-to-image.
By tracking the evolution of the resulting speckle brightness as a
function of the probe phase, one is able to determine the true
phase $\varphi_0$ of the original speckle.
Once the speckle phase is determined, a correction of opposite phase
$\varphi_0+\pi$ and of amplitude $g \times a_0$ (with $g\sim0.1$
the loop gain), is permanently applied before the loop is allowed
to hunt for other speckles.
A minimum of four probes with phase 0, $\pi/2$, $\pi$ and $3\pi/2$
is required to unambiguously estimate a long-lived quasi-static
speckle phase. Unfortunately, the 10-ms exposure time imposed by
the characteristics of the detector happens to be of the order of
the time scale of the AO188 wavefront correction. To make the
speckle nulling loop robust against fast fluctuations associated
to the dynamic speckle component, the algorithm uses 30 probes of
phase uniformly varying between 0 and $2\pi$.

After the speckle nulling loop has converged, a mirror located in a
collimated beam after the coronagraph (cf\ Fig.\ \ref{f:layout}) is
translated out of the optical path, so as to direct the light toward
the HiCIAO imager.

Given the chosen number of probes, frame rate, and processing time, we
get one iteration completed in 4 seconds. We increase the 
efficiency of the loop by simultaneously probing as many speckles as
possible. The image-to-image variance due to the dynamic aberrations
and the poor sensitivity of the camera however set a limit to the
total number of speckles that can be probed. In practice, the
algorithm tested on-sky was able to simultaneously probe an average of
five of the brightest speckles in the field.

\subsection{Speckle nulling data}

\begin{figure*}
\plottwo{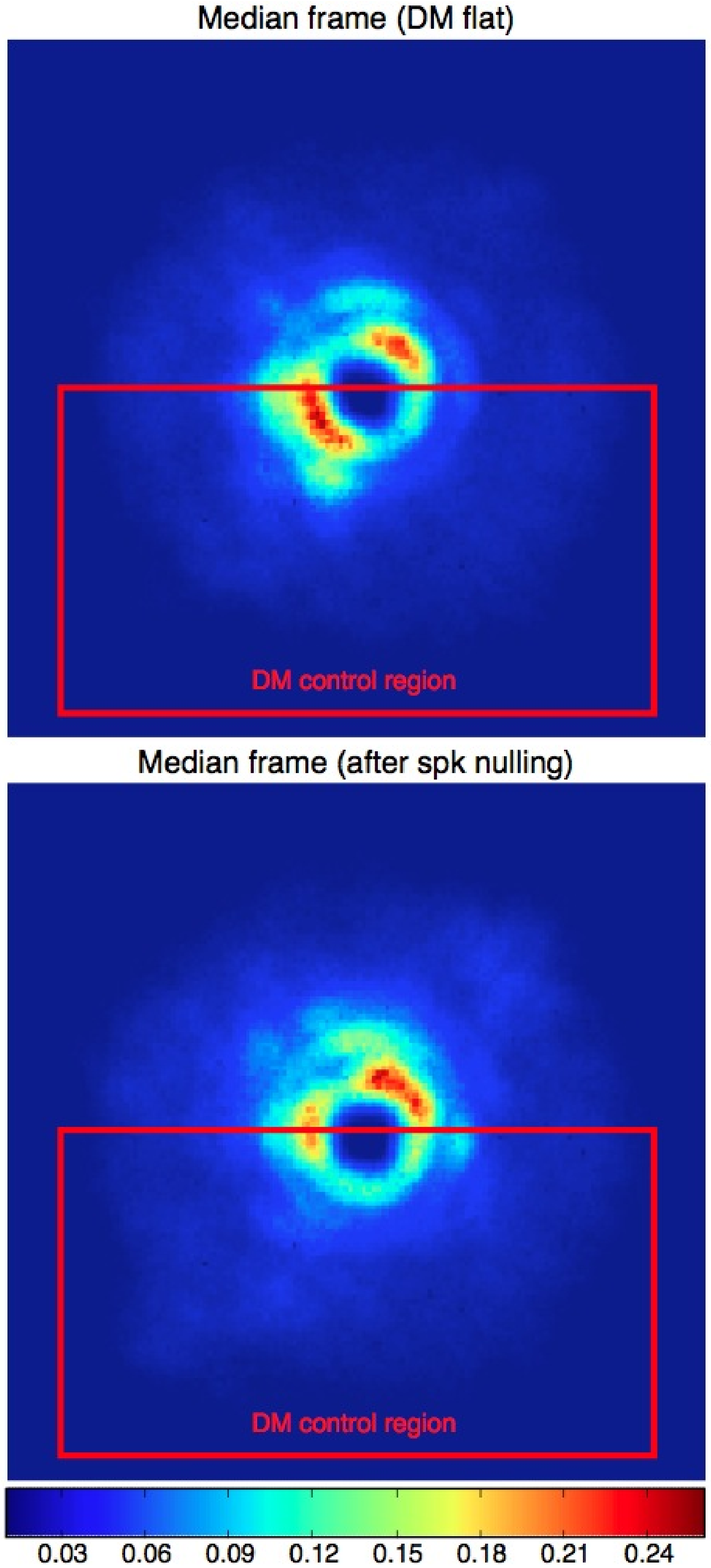}{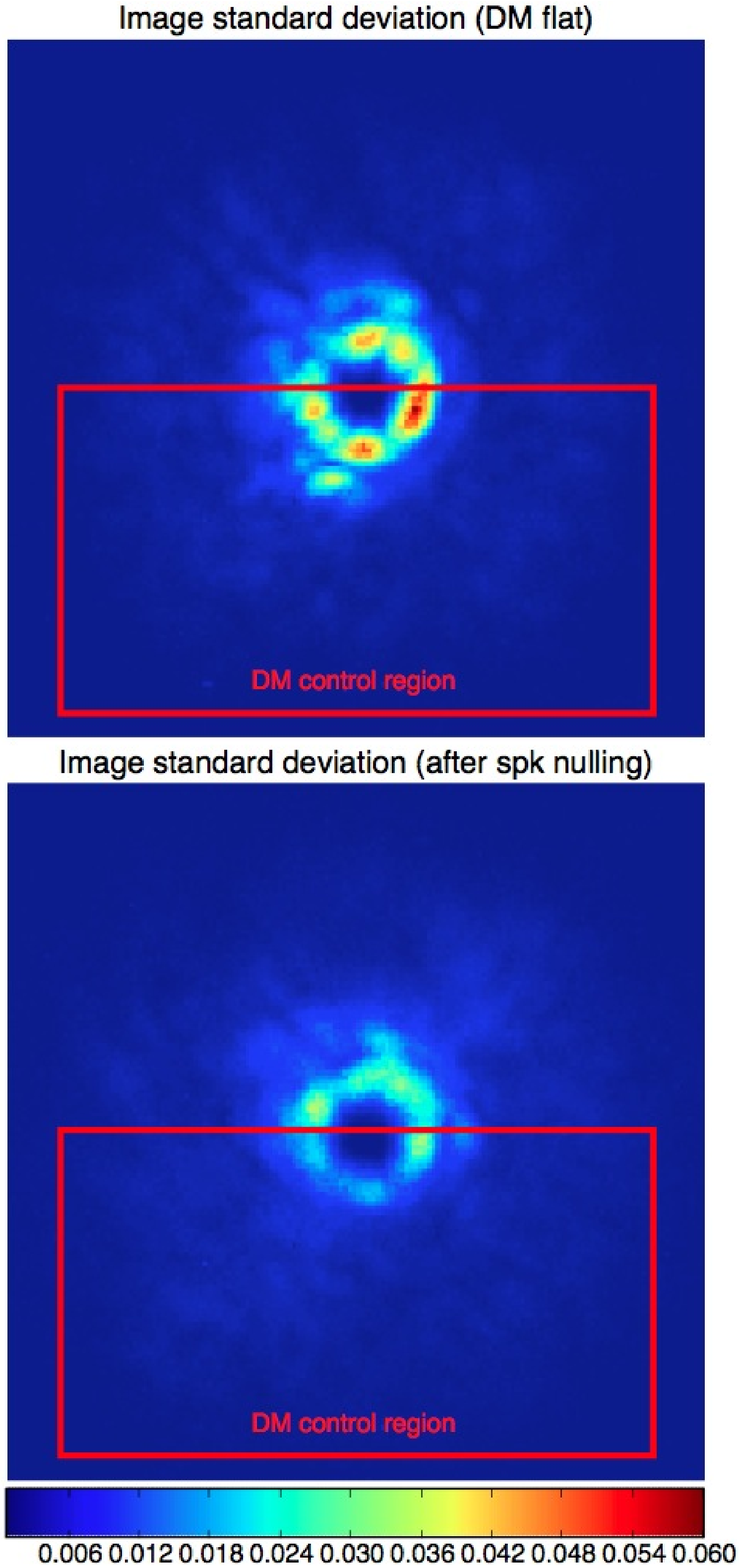}
\caption{
Statistical properties (median and standard deviation) of
coronagraphic images acquired by HiCIAO for two SCExAO DM shapes. The
top row is for a flat DM shape, that maximizes the Strehl ratio. The
bottom row is for a DM shape resulting from a speckle nulling loop.
The colorbar scale used for the images on the left (median) is
calibrated in contrast. 
}
\label{f:res} 
\end{figure*}

Twelve minutes of the speckle nulling loop running were sufficient to
be able to visually observe a darkening of one side of the field of
view.
To test the efficiency of the wavefront modulation resulting from the
speckle nulling loop, several series of images were acquired with
HiCIAO, alternating between a ``flat'' and an ``optimized'' SCExAO DM
shape.
The flat DM shape resulted from an optimization routine maximizing
the Strehl in the non-coronagraphic image, using Zernike modes:
astigmatism, focus, coma, spherical, trefoil and quadrifoil on an
internal calibration source.
The optimized shape was the result of speckle nulling corrections
applied in addition to the flat.
The complete acquisition sequence consisted of four such cycles (to
make the analysis robust against slow trend effects), resulting in a
total of twenty 5-second exposure frames per DM shape.
These HiCIAO images were dark-subtracted and flat-fielded, using
standard data reduction procedures established for the reduction of
the SEEDS data, assembled into two data cubes.

The images gathered in Fig. \ref{f:res} summarize the
statistical properties of these two datacubes, looking at the median
(left column) and the standard deviation (right column), for the
flat (top row) as well as the optimized DM shape (bottom row), using
common colorbars.
For each figure, the extent of the control region is highlighted by a
red box, which for this experiment, is located in the lower part of
the HiCIAO image.

The combination of exposure time (5 seconds) and neutral density
(ND0.1) for HiCIAO was chosen to acquire non-saturated images in the
H-broadband filter. The low and fast varying seeing experienced during
the acquisition and the lack of a yet to come high-order close-loop
adaptive correction results in exposures that exhibit no obvious
individual speckles in the field, a major difference from what is
already achieved by XAO equiped instruments (see for instance Fig. 1
of \citet{2013ApJ...768...24O}).
The first bright ring, located at a radius of 2.3 $\lambda/D$
is the dominating feature in the images. Note that this radius is very
close to the quoted IWA of SCExAO (2.2 $\lambda/D$), defined as the
angular separation for which the coronagraph throughput reaches 50\%
\citep{2012PASP..124.1288M}, and imposed by the focal plane mask.
The standard deviation of the DM flat datacube shows that most of the
speckle variance is localised onto the diffraction rings.

The speckle nulling algorithm unsurprisingly focused its efforts onto
this area of the image. With the optimized DM shape (bottom images in
Fig.\ \ref{f:res}), the median flux for the part of the diffraction
ring overlapping with the control region is reduced. The improvement
in regards to the median is moderate as the brightest parts of the
ring are attenuated by a factor 2.

\subsection{Benefit for high contrast imaging}

The detection of faint structures in a speckle-noise dominated problem
is usually achieved by some form of PSF subtraction. The PSF can be
obtained from more or less sophisticated methods: from the direct
observation of a reference star to a synthetic PSF obtained after ADI.
PSF subtraction is obviously appropriate to calibrate static features
of the PSF. It is also good for fast-varying speckles that average
down quickly to a smooth halo.

PSF subtraction however fails to calibrate speckles characterized by a
timescale of a few minutes which manifest by residual variance in PSF
subtracted images: the speckle noise.
The top row of Fig. \ref{f:res} illustrates this phenomenon and shows
that over the course of the twelve minutes covered by the experiment,
on the most prominent image features, the fluctuations are on the
order of 25-50\% of the local median value, and that the overall
structure of the standard deviation image follows that of the median:
static aberrations do appear to mix with and amplify the local speckle
noise, resulting in speckles pinned on the static diffraction pattern
\citep{2004ApJ...612L..85A}.

In actively suppressing the static features of the coronagraphic PSF
(cf. bottom row of Fig. \ref{f:res}), speckle nulling achieves two
goals: it removes slow and static speckles (measurable in comparing
the flat and optimized median frames) and reduces the speckle noise
(measurable in comparing the standard deviation frames). The PSF mean
level over the region corrected by speckle nulling is not only lower
by a factor of 2, it is also more stable, with a standard deviation
reduced by a factor 3.

\section{Conclusion}

This paper presents the first demonstration of speckle nulling
performed on-sky with partial AO correction. In the context of the
SCExAO project, using a PIAA coronagraph and a deformable mirror
to modulate the focal plane, the results presented here show that it
is indeed possible to complement ADI and its variants at small angular
separations, even in the absence of a high strehl AO system, and
therefore in the presence of a brighter dynamic speckle halo.

Speckle nulling offers improvement at two levels: (1) it removes slow
speckles, resulting in images with enhanced contrast before
processing, and (2) it reduces the speckle noise amplification by the
static diffraction, resulting in enhanced post-processing detection
limits. Moreover, it enables this in a regime of angular separation
that cannot be addressed by ADI, which remains the reference technique
\citep{2013ApJ...764..183B} for the calibration of ground based AO
PSF. In spite of unfavorable observing conditions, the control loop
managed to remain stable and converged to improve the statistical
properties of the coronagraphic image.

While consistent, the reported improvement remains modest, and this
can in part be attributed to the choices of integration time,
constrained by the noise properties of the detector used during the
speckle nulling loop.
Improving the sensitivity and the dynamical range of the image by
using a cooled detector will increase the flexibility of the
technique, and allow a more efficient use of the observing time.
The real performance improvement will however come from the of an
actual close-loop XAO system that will improve the raw Strehl of
individual images.

Advanced focal plane based wavefront techniques such as electric field
conjugation \citep{2006PhDT........47G,2006ApJ...638..488B} have been
used extensively on laboratory high contrast test-beds over the world,
including with a PIAA coronagraph like the one used on SCExAO
\citep{2010PASP..122...71G} providing access to very high contrast
($10^{-7}$ and beyond) detection limits in a stable environment.
The fact that even in difficult observing conditions, a speckle
nulling driven wavefront control algorithm, originally envisioned for
stable test-benches, remains operational, is very encouraging. 
This indeed suggests that with only a few adaptations, all of the high
contrast laboratory work that has so far been geared toward space
borne applications is relevant to ground based observations, even with
partial AO correction.


\end{document}